%% file: ae-sagt09-long.tex
\documentclass{llncs}
\usepackage{amsmath}
\usepackage{amsfonts}
\usepackage{amssymb}

\usepackage{latexsym}

\include{ecmds}
\usepackage{srcltx}
\usepackage{srctex}

\newcommand{\Proof}{\NI
                    {\bf Proof.}\ }
\def\smallromani{\renewcommand{\theenumi}{\roman{enumi}}
\renewcommand{\labelenumi}{(\theenumi)}}

\title{Sequential pivotal mechanisms for public project problems}

 \author{Krzysztof R. Apt \inst{1,2}
\and \mbox{Arantza Est\'{e}vez-Fern\'{a}ndez} \inst{3}
 }
 \institute{Centrum Wiskunde \& Informatica (CWI), Amsterdam
\and University of Amsterdam
\and Dept. of Econometrics and Operations Research, VU University Amsterdam
}

\date{}
\begin{document}
\maketitle

\begin{abstract}
  It is well-known that for several natural decision problems no
  budget balanced Groves mechanisms exist. This has motivated recent
  research on designing variants of feasible Groves mechanisms (termed
  as `redistribution of VCG (Vickrey-Clarke-Groves) payments') that
  generate reduced deficit.  With this in mind, we study sequential
  mechanisms and consider optimal strategies that could reduce the
  deficit resulting under the simultaneous mechanism. We show that
  such strategies exist for the sequential pivotal mechanism of the
  well-known public project problem.  We also exhibit an optimal
  strategy with the property that a maximal social welfare is
  generated when each player follows it. Finally, we show that these
  strategies can be achieved by an implementation in Nash equilibrium.
All proofs can be found in the full version posted in 
Computing Research Repository (CoRR), \url{http://arxiv.org/abs/0810.1383}.
\end{abstract}

\section{Introduction}
\label{sec:intro}
\subsection{Motivation}

Mechanism design is concerned with designing non-coope\-rative games
in which the participating rational players achieve the desired social
outcome by reporting their types.  Among the most commonly studied
mechanisms are the ones in the Groves family that are based on
transfer payments (taxes).  For the case of efficient decision
functions they are \emph{incentive compatible}, i.e.,
they achieve truth-telling in dominant strategies.
The special case called \emph{pivotal mechanism} (sometimes also
called \emph{VCG} (Vickrey-Clarke-Groves) mechanism) is additionally
\emph{pay only} (i.e., each player needs to pay a tax) and hence
\emph{feasible} (i.e., the generated deficit is negative or zero).

It is well-known that for several problems incentive compatible
mechanisms cannot achieve \emph{budget balance} (which states that the generated
deficit is zero), see, e.g., Chapter 23 of \cite{MWG95}.  This has
motivated recent research in designing appropriate instances of Groves
mechanisms that generate a reduced deficit (or equivalently higher
social welfare). These modifications are termed as `redistribution of
VCG payments'. In fact, they are variants of feasible Groves
mechanisms.  Notably, \cite{Cav06} and \cite{GC07} showed that a
deficit reduction is possible for the case of Vickrey auctions
concerned with multiple units of a single good.  On the other hand,
\cite{ACGM08} recently showed that no such deficit reduction is
possible in the well-known case of the pivotal mechanism for the
public project problem.

This research direction motivates our study of sequential Groves
mechanisms, in particular sequential pivotal mechanism, in which
players move sequentially. We face then a new situation since each
player knows the types reported by the previous players.  Sequential
Groves mechanisms apply to a realistic situation in which there is no
central authority that computes and imposes taxes and where the
players move in a randomly chosen order.

\subsection{Contributions}

We show here that natural strategies exist in the sequential pivotal
mechanism for the public project problem that generate larger social
welfare than truth-telling.  We also exhibit a strategy such that the
social welfare is maximized when each player follows it. Finally, we
show that the resulting sequential mechanisms yield an implementation
in Nash equilibrium. Moreover, the vector of the latter strategies is
also Pareto optimal.

To properly describe the nature of the introduced strategies we
consider two concepts.  An \oldbfe{optimal} strategy guarantees a
player the maximum utility under the assumption that he moves
simultaneously with the players who follow him.  It also
guarantess the player at least the same utility as truth-telling,
under the assumption that the other players are truth-telling.
% player at least the same utility as truth-telling, under the
% assumption that the other players are truth-telling.  In turn, a
In turn, a \oldbfe{socially optimal} strategy yields the maximal
social welfare among all optimal strategies.
% under the assumption that the other players are truth-telling. (In
% fact, both concepts are slightly more general.)

These concepts allow us to analyze altruistic behaviour of the players
in the framework of sequential pivotal mechanism. By altruistic
behaviour we mean that the players do not only care about their own
utility, but also about the utility of the others.

\subsection{Related work}

Ever since the seminal paper of \cite{Cla71} mechanism design for
public goods has received a huge attention in the literature.
We mention here only some representative papers the results of which
provide an appropriate background for our work.

Both the continuous and discrete case of public goods have been studied.
The former situation has been in particular considered in \cite{GL77},
where a taxation scheme has been proposed which leads to a Pareto optimal
solution that can be realized in a Nash equilibrium.
Sequential mechanism design for public good problems
has been considered in \cite{CR85}, where a ``Stackelberg'' mechanism
was proposed that combines optimal Bayes strategies with dominant
strategies.

% In the related context of the implementation theory
% sequential implementation has been considered in \cite{MR88}, where an
% implementation by means of a subgame perfect equilibrium has been
% studied.

Here we study the discrete case.  The situation when the decision is
binary (whether to realize a public project or not) has been studied
in \cite{JM92}, where balanced but not incentive compatible sequential
mechanisms have been proposed. These mechanisms can be realized in an
undominated Nash equilibrium and in subgame perfect equilibrium.  Many
aspects of incentive compatible mechanisms for public goods have been
surveyed in \cite{Che08}.

The consequences of sequentiality have also been studied in the
context of private contributions to public goods and in voting theory.
In particular, \cite{Var94} has studied the behavior of players
depending on the position in which they have to take a decision and
\cite{DP00} has explored the relationship between simultaneous and
sequential voting games.  More recently, \cite{LPRVW07} has studied
the problem of determining the winner in elections in which the voting
takes place sequentially.

Our focus on maximizing social welfare is related to research on
altruistic behaviour of the player. This subject has been studied in a
number of papers in game theory, most recently in \cite{MM07}, where
several references to earlier literature on this subject can be found.
Finally, in a recent work, \cite{AM09}, we carried out an analogous
analysis for two feasible Groves mechanisms used for single item
auctions: the Vickrey auction and the Bailey-Cavallo mechanism.

\subsection{Plan of the paper}

The paper is organized as follows.  In the next section we recall
Groves mechanisms and the pivotal mechanism by focusing on decision
problems.  Then, in Section \ref{sec:sequential1}, we introduce
sequential decision problems, in particular sequential Groves
mechanisms.

In the remaining sections we study the sequential pivotal mechanism
for the public project problem.  In Section
\ref{sec:example1} we exhibit an optimal strategy that in a limited sense simultaneously
maximizes players' final utilities and another optimal
strategy that maximizes the social welfare among all vectors of optimal strategies.
Finally, in Section \ref{sec:nash}, we
clarify the status of the optimal strategies
introduced in Section \ref{sec:example1} by showing that their vector
is a Nash equilibrium w.r.t.~appropriately defined preference
relations on the strategy vectors, and by providing a corresponding
revelation-type result.  We conclude by mentioning some open problems
in Section \ref{sec:conc}.

\section{Preliminaries}\label{sec:prelim}

We briefly recall the family of Groves mechanisms here. In this
section we follow \cite{Jac03}.  Let $D$ be a set of
\oldbfe{decisions}, $\{1, \LL, n\}$ be the set of players with $n \geq
2$, and for each player $i$ let $\Theta_i$ be a set of his \oldbfe{types} and
$v_i : D \times \Theta_i \myra \mathbb{R}$ be his (\oldbfe{initial})
\oldbfe{utility function}.

A \oldbfe{decision rule} is a function $f: \Theta \myra D$, where
$\Theta := \Theta_1 \times \cdots \times \Theta_n$.  
It is called 
% \oldbfe{strategy-proof}
% if for all $\theta \in \Theta$, $i \in \C{1,\LL,n}$ and
% $\theta'_i \in \Theta$
% \[
% v_i(f(\theta_i, \theta_{-i}), \theta_i) \geq
% v_i(f(\theta'_i, \theta_{-i}), \theta_i),
% \]
\oldbfe{efficient} if for all $\theta \in
\Theta$ and $d' \in D$
\[
\sum_{i = 1}^{n} v_i(f(\theta), \theta_i) \geq \sum_{i = 1}^{n} v_i(d', \theta_i).
\]

We call the tuple $(D, \Theta_1, \LL, \Theta_n, v_1, \LL, v_n, f)$ a
\oldbfe{decision problem}.  
% Given a decision problem, one is
% interested in the following sequence of events:

% \begin{enumerate} \smallromani
% \item each player $i$ receives a type $\theta_i \in \Theta_i$,

% \item each player $i$ announces a type $\theta'_i  \in \Theta_i$ to the central planner;
% this yields a joint type $\theta' := (\theta'_1, \LL, \theta'_n)$,
% \label{item:2}

% \item the central planner takes a decision $d := f(\theta')$,
% and communicates it to each player,
% \label{item:3}

% \item the resulting utility for player $i$ is then
% $v_i(d, \theta_i)$.
% \end{enumerate}

Recall that a \oldbfe{direct mechanism} is obtained by
transforming the initial decision problem
$(D, \Theta_1, \LL, \Theta_n, v_1,
\LL, v_n, f)$ as follows:
\begin{itemize}

\item the set of decisions is
$
D \times \mathbb{R}^n,
$

\item the decision rule is a function
$
(f,t): \Theta \myra D \times \mathbb{R}^n,
$
where $
t: \Theta \myra \mathbb{R}^n
$
and
$
(f,t)(\theta) := (f(\theta), t(\theta)),
$

\item each \oldbfe{final utility function} for player $i$ is a function $u_i: D \times \mathbb{R}^n \times \Theta_i \myra \mathbb{R}$
defined by
$
u_i(d,t_1, \LL, t_n, \theta_i) :=  v_i(d, \theta_i) + t_i.
$
\end{itemize}
We call then $\sum_{i = 1}^{n} u_i((f,t)(\theta), \theta_i)$
the corresponding \oldbfe{social welfare} and refer to $t$ as the \oldbfe{tax function}.

A direct mechanism with tax function $t$ is called
  \begin{itemize}

  \item (\oldbfe{dominant strategy}) \oldbfe{incentive compatible}
% if the decision rule $(f,t)$ is strategy-proof,
if for all $\theta \in \Theta$, $i \in \C{1,\LL,n}$ and
$\theta'_i \in \Theta_i$
\[
u_i((f,t)(\theta_i, \theta_{-i}), \theta_i) \geq
u_i((f,t)(\theta'_i, \theta_{-i}), \theta_i),
\]

  \item \oldbfe{budget balanced} if $\sum_{i = 1}^{n} t_i(\theta) = 0$ for all $\theta$,

  \item \oldbfe{feasible} if $\sum_{i = 1}^{n} t_i(\theta) \leq 0$ for all $\theta$,

  \item \oldbfe{pay only} if $t_i(\theta) \leq 0$ for all $\theta$ and all $i \in \{1, \LL, n\}$.

  \end{itemize}

Each \oldbfe{Groves mechanism} is obtained by using the tax function
$
t := (t_1, \LL, t_n),
$
where\footnote{Here and below $\sum_{j\not=i}$ is a
  shorthand for the summation over all $j \in \{1,\LL,n\}, \ j
  \not=i$.} for all $i \in \C{1, \LL, n}$
\[
t_i(\theta) :=  \sum_{j \neq i} v_j(f(\theta), \theta_j) + h_i(\theta_{-i}),
\]
with $h_i: \Theta_{-i} \myra \mathbb{R}$ an arbitrary function.

Finally, we recall the following crucial result.
\II

\NI
\textbf{Groves Theorem}
Consider a decision problem with an efficient
decision rule $f$. Then each Groves mechanism is incentive compatible.
\II

A special case of Groves mechanism is the \oldbfe{pivotal mechanism}, which is a pay only
mechanism obtained by
$
h_i(\theta_{-i}) := - \max_{d \in D} \sum_{j \neq i} v_j(d, \theta_j).
$

Direct mechanisms for a given decision problem can be compared
w.r.t.~the social welfare they entail. More precisely, given
a decision problem 
\[
(D, \Theta_1, \LL, \Theta_n, v_1, \LL, v_n, f)
\] 
and
direct mechanisms (determined by the sequences of tax
functions) $t$ and $t'$ we say that $t'$ \oldbfe{welfare dominates}
$t$ if

\begin{itemize}

\item for all $\theta \in \Theta$
\[
\sum_{i = 1}^{n} u_i((f,t)(\theta), \theta_i) \leq \sum_{i = 1}^{n} u_i((f,t')(\theta), \theta_i),
\]

\item for some $\theta \in \Theta$
\[
\sum_{i = 1}^{n} u_i((f,t)(\theta), \theta_i) < \sum_{i = 1}^{n} u_i((f,t')(\theta), \theta_i).
\]
\end{itemize}

In this paper we analyze the following well-known decision problem, originally due to \cite{Cla71},
and extensively discussed in the economic literature, see, e.g. \cite{MWG95,Jac03}.

\paragraph{Public project problem}
\mbox{} \\
\NI
Consider $
(D, \Theta_1, \LL, \Theta_n, v_1, \LL, v_n, f),
$
where

\begin{itemize}
\item $D = \{0, 1\}$
(reflecting whether a project is cancelled or takes place),

\item for all $i \in \{1, \LL, n\}$,
$\Theta_i = [0,c]$, where $c > 0$,

\item for all $i \in \{1, \LL, n\}$, $v_i(d, \theta_i) := d (\theta_i - \frac{c}{n})$,

\item $
        f(\theta) :=
        \left\{
        \begin{array}{l@{\extracolsep{3mm}}l}
        1    & \mathrm{if}\  \sum_{i = 1}^{n} \theta_i \geq c \\
        0       & \mathrm{otherwise}
        \end{array}
        \right.
$
\end{itemize}

In this setting $c$ is the cost of the project, $\frac{c}{n}$ is the cost
share of the project for each player, and $\theta_i$ is the value
of the project for player $i$. Note that the decision rule $f$ is efficient since
$\sum_{i = 1}^{n} v_i(d, \theta_i) = d(\sum_{i = 1}^{n} \theta_i - c)$.

% It is well-known that in the case of the public project problem for no $c > 0, n \geq 2$
% and $\Theta$ pivotal mechanism is budget balanced.
% Indeed, take $\epsilon \in (0, \frac{c}{n})$ and
% consider $\theta_1 = \frac{c}{n} + \epsilon$, $\theta_i = \frac{c}{n}$ for $i \in \{2, \LL, n-1\}$ and $\theta_n = 0$.
% Then $f(\theta) = 0$ and $\sum_{j \neq n} v_j(1, \theta_j) \geq \sum_{j \neq n} v_j(0, \theta_j)$.
% Consequently $t_n(\theta) = - \epsilon$, while for $i \in \{1, \LL, n-1\}$, $t_i(\theta) \leq 0$.
% So $\sum_{i = 1}^{n} t_i(\theta) < 0$.

It is well-known that for no $n \geq 2$ and $c > 0$ an incentive
compatible direct mechanism for the public project problem exists that
is budget balanced, see, e.g.~\cite[ page 861-862]{MWG95}.  It is then
natural to search for incentive compatible direct mechanisms that
generate a smaller deficit than the one obtained by the pivotal
mechanism.  However, the following optimality result concerning the
pivotal mechanism, recently established in \cite{ACGM08}, dashed 
hope.

\begin{theorem} \label{thm:opt}
  In the public project problem there exists no feasible incentive
  compatible direct mechanism that welfare dominates the pivotal
  mechanism.  
\HB
\end{theorem}

Our aim is to show that when the original setting of the public
project problem is changed to one where all players announce their
types sequentially in a random order, then the deficit can be reduced.

\section{Sequential decision problems}\label{sec:sequential1}

In this section we introduce sequential decision problems. For
notational simplicity, and without loss of generality, we assume
that players sequentially report their types in the order $1, \LL,
n$. To capture this type of situations, given a decision problem
${\cal D} := (D, \Theta_1, \LL, \Theta_n, v_1, \LL, v_n, f)$, we
% consider a modified sequence of events in which events
% (\ref{item:2}) and (\ref{item:3}) of Section \ref{sec:prelim} are
% replaced by:
% \medskip
%
% \NI 
%\ (ii)$^{\prime}$ successively stages $1, \LL, n$ take place, where in
assume that successively stages $1, \LL, n$ take place, where in
stage $i$   player $i$ announces  a type $\theta'_i$ to the \emph{other players}.
After stage $n$ this yields a joint type $\theta' := (\theta'_1,
\LL, \theta'_n)$. Then each player takes the decision $d := f(\theta')$.
\NI 
%\ (iii)$^{\prime}$ each player takes the decision $d := f(\theta')$.

%\medskip

We call the resulting situation a \oldbfe{sequential decision
problem} or more specifically, a \oldbfe{sequential version of
${\cal D}$}. Note that in a sequential decision problem a central
planner may not exist and decisions may be taken by the players
themselves. Each player $i$ \emph{knows} the types announced by
players $1, \LL, i-1$, so that he can use this information to decide
which type to announce. To properly describe this situation we need
to specify what is a strategy in this setting.  A \oldbfe{strategy}
of player $i$ in the sequential version of ${\cal D}$ is a function
\[
s_i: \Theta_1 \times \LL \times \Theta_i \myra \Theta_i.
\]
In this context \oldbfe{truth-telling}, as a strategy, is represented by the
projection function $\pi_{i}(\cdot)$, defined by
$\pi_i(\theta_1,\LL,\theta_i) :=\theta_{i}$.

From now on, we consider a direct mechanism
\[
{\cal D} := (D \times \mathbb{R}^n, \Theta_1, \LL, \Theta_n, u_1, \LL, u_n, (f,t))
\] 
and mainly focus on Groves mechanisms.

We assume that in the considered sequential decision problem each
player uses a strategy $s_i(\cdot)$ to select the type he will
announce.  We say that strategy $s_i(\cdot)$ of player $i$ is
\oldbfe{optimal} in the sequential version of ${\cal D}$ if for all
$\theta \in \Theta$ and $\theta'_i \in \Theta_i$
\[
u_i((f,t)(s_i(\theta_1,\LL, \theta_i), \theta_{-i}), \theta_i) \geq
u_i((f,t)(\theta'_i, \theta_{-i}), \theta_i).
\]
% Here as before, $\theta_i$ is the type that player $i$ has received,
% while $\theta_{-i}$ is the vector of announced types of the other players.

Call a strategy of player $j$ \oldbfe{memoryless} if it does not depend on
the types of players $1, \LL, j-1$. Then a strategy $s_i(\cdot)$ of
player $i$ is optimal if for all $\theta \in \Theta$ it yields a best
response to all joint strategies of players $j \neq i$ under the
assumption that players $i+1, \LL, n$ use memoryless strategies or move
jointly with player $i$.  In particular, an optimal strategy is a best
response to the truth-telling by players $j \neq i$.
% By choosing truth-telling as the
% strategies of players $j \neq i$ we see that each dominant strategy is
% optimal. For player $n$ the concepts of dominant and
% optimal strategies coincide.

A particular case of sequential decision problems are sequential
Groves mechanisms.  The following direct consequence of Groves Theorem
provides us with a simple method of determining whether a strategy is
optimal in such a mechanism.

\begin{lemma} \label{lem:vcg1}
Let $(D, \Theta_1, \LL, \Theta_n, v_1, \LL, v_n, f)$ be a decision
problem with efficient decision rule $f$. Suppose that $s_i(\cdot)$
is a strategy for player $i$ such that for all $\theta \in \Theta$,
$f(s_i(\theta_1,\LL,\theta_i), \theta_{-i})=f(\theta_i,
\theta_{-i})$. Then $s_i(\cdot)$ is optimal in each sequential
Groves mechanism $(D \times \mathbb{R}^n, \Theta_1, \LL, \Theta_n,
u_1, \LL, u_n, (f,t))$.
\HB
\end{lemma}

\II

In particular, when the decision rule is efficient, the truth-telling
strategy $\pi_i(\cdot)$ is optimal in each sequential
Groves mechanism.

We are interested in maximizing the social welfare. This motivates
the following notion. We say that
strategy $s_i(\cdot)$ of player $i$ is \oldbfe{socially optimal} in
the sequential version of ${\cal D}$ if it is optimal and for all
optimal strategies $s'_i(\cdot)$ of player $i$ and all $\theta \in
\Theta$ we have
\[
 \begin{array}{l}
\sum_{j=1}^{n} u_j((f,t)(s_i(\theta_1,\LL, \theta_i), \theta_{-i}),
\theta_j) \geq \sum_{j=1}^{n} u_j((f,t)(s'_i(\theta_1,\LL, \theta_i),
\theta_{-i}), \theta_j).
\end{array}
\]

Hence a socially optimal strategy of player $i$ yields the maximal
social welfare among all optimal strategies, under the assumption that
players $i+1, \LL, n$ use memoryless strategies or move jointly.

Consider now a sequential version of a given direct mechanism 
\[(D \times \mathbb{R}^n, \Theta_1, \LL, \Theta_n, u_1, \LL, u_n, (f,t))
\]
and assume that each player $i$ receives a type $\theta_i \in
\Theta_i$ and follows a strategy $s_i(\cdot)$. The resulting social
welfare is then
\[
SW(\theta, s(\cdot)) := \sum_{j=1}^{n} u_j((f,t)([s(\cdot),
\theta]), \theta_j),
\]
where $s(\cdot) := (s_1(\cdot), \LL, s_n(\cdot))$ and $[s(\cdot),
\theta]$ is defined inductively by $[s(\cdot), \theta]_1 :=
s_{1}(\theta_1)$ and $[s(\cdot), \theta]_{i+1} := s_{i+1}([s(\cdot),
\theta]_1, \LL, [s(\cdot), \theta]_i, \theta_{i+1})$.

In general, if player $i$ assumes that he moves jointly with players $i+1, \LL, n$
he will choose an optimal strategy. And if additionally
he wants to maximize the social welfare, he will choose a socially
optimal strategy (if it exists).  In the next section we shall see
that for the public project problem a sequence of socially optimal
strategies can be found for which the resulting social welfare is
always maximal.  In general, we only have the following limited
result.

\begin{lemma} \label{lem:optn}
Consider a direct mechanism $(D \times \mathbb{R}^n,
\Theta_1, \LL, \Theta_n, u_1, \LL, u_n, (f,t))$ and let $s_n(\cdot)$
be a socially optimal strategy for player $n$. Then
\[
SW(\theta, (s'_{-n}(\cdot), s_n(\cdot))) \geq SW(\theta, s'(\cdot))
\]
for all $\theta \in \Theta$ and vectors $s'(\cdot)$ of optimal
players' strategies.
\end{lemma}

\Proof
Directly by the definition of a socially optimal strategy.
\HB

\section{Public project problem}
\label{sec:example1}

In what follows, we focus on the special case of sequential pivotal
mechanisms for the public project problem.  First, the following
theorem gives an optimal strategy for player $i$ that may differ from
truth-telling.  Part $(ii)$ shows that, under certain natural
conditions, this strategy simultaneously maximizes the final utility
of every other player.

\begin{theorem} \label{thm:dom1}
Let ${\cal D}$ be a public project problem. Let
\[
s_i(\theta_1, \LL, \theta_i) :=
 \left\{
        \begin{array}{l@{\extracolsep{3mm}}l}
        \theta_i    & \mathrm{if}\  \sum_{j=1}^{i}\theta_j < c\mathrm{\ and \ } i<n\mbox{,} \\
        0  & \mathrm{if}\  \sum_{j=1}^{i}\theta_j < c\mathrm{\ and \ } i=n\mbox{,} \\
        c       & \mathrm{if \ }\ \sum_{j=1}^{i}\theta_j \geq c
        \end{array}
        \right.
\]
be strategy for player $i$. Then

\begin{enumerate} \smallromani
\item $s_i(\cdot)$ is optimal for player $i$ in the sequential pivotal
  mechanism,

\item for all $\theta \in \Theta$ and $\theta'_i \in \Theta_i$ such that 
$s_i(\theta_1,\LL,\theta_i) \neq \theta_i$ and
$f(\theta'_i, \theta_{-i}) = f(\theta_i, \theta_{-i})$ we have
for all $j \neq i$
\[
u_j((f,t)(s_i(\theta_1,\LL, \theta_i), \theta_{-i}), \theta_j) \geq
u_j((f,t)(\theta'_i, \theta_{-i}), \theta_j).
\]

\end{enumerate}
\end{theorem}

%We relegate the proof to the Appendix. 
In part $(ii)$ $\theta_{-i}$ are the types submitted
by players $j \neq i$ and $\theta_i$ is the type
received by player $i$.  So part $(ii)$ states that if
strategy $s_i(\cdot)$ of player $i$ deviates from truth-telling
($s_i(\theta_1,\LL,\theta_i) \neq \theta_i$) and the players who
follow $i$ use memoryless strategies (so in particular, the types they
submit do not depend on the type submitted by player $i$), then player
$i$ simultaneously maximizes the final utility of the other players
(and hence the social welfare). This happens under the assumption that
player $i$ submits a type that does not alter the decision to be taken.

% Let us focus now on the pivotal mechanism.
% Recall now that if player
% $i$ submits a type $\theta'_i$ and the other submitted types are
% $\theta_{-i}$, then each player $j \neq i$ pays the tax $|t_j(\theta'_i,
% \theta_{-i})|$, where
% \[
% \begin{array}{l}
% t_j(\theta'_i,\theta_{-i}) :=  \\
% \quad \quad \sum_{k \neq i,j} v_k(f(\theta'_i,\theta_{-i}), \theta_j) + v_i(f(\theta'_i,\theta_{-i}), \theta'_i) - \\
% \quad \quad \max_{d \in D} (\sum_{k \neq i,j} v_k(d, \theta_j) + v_i(d, \theta'_i)).
% \end{array}
% \]

% \[
% t_j(\theta'_i,\theta_{-i}) := \sum_{k \neq i,j} v_k(f(\theta'_i,\theta_{-i}), \theta_j) + v_i(f(\theta'_i,\theta_{-i}), \theta'_i) -
% \max_{d \in D} (\sum_{k \neq i,j} v_k(d, \theta_j) + v_i(d, \theta'_i)).
% \]
%

% We have $t_j(\theta'_i,\theta_{-i}) \leq 0$, so in the sequential pivotal mechanism
% player $i$, when using
% Lemma \ref{lem:vcg1} to minimize player's $j$ tax, solves the
% following maximization problem:
% \[
%  \begin{array}{l}
% \mbox{maximize $t_j(\theta'_i, \theta_{-i})$ subject to $\theta'_i \in \Theta_i$ and $f(\theta'_i, \theta_{-i}) = f(\theta_i, \theta_{-i})$.}
% \end{array}
% \]

% \begin{equation}\label{eqclarke}
%  \begin{array}{l}
% \mbox{maximize $t_j(\theta'_i, \theta_{-i})$ subject to}\\
% \mbox{$\theta'_i \in \Theta_i$ and $f(\theta'_i, \theta_{-i}) = f(\theta_i, \theta_{-i})$.}
% \end{array}
% \end{equation}

When each player follows strategy $s_i(\cdot)$, always the same
decision is taken as when each player is truthful, independently on
the players' order. Additionally, by part $(ii)$ of Theorem
\ref{thm:dom1} with $\theta'_i = \theta_i$, social welfare weakly
increases.  The following example shows that sometimes a strictly
larger social welfare can be achieved.

\begin{example} \label{exa:tables}
  Assume that $c = 300$ and that there are three players, A, B and C.
  Table \ref{tab:si1} illustrates the situation in the case of pivotal
  mechanism.  In Table \ref{tab:si2} we assume that the players submit
  their types in the order A, B, C. Here the social welfare increases
  from $-20$ to $-10$.
\HB
\end{example}

\begin{table}[t]%[htbp]
\begin{center}
\begin{tabular}{|c|c|c|c|c|}
\hline
player & type & submitted type & tax & $u_i$ \\\hline
A & $110$  & $110$ & $-10$ & $0$\\\hline
B & $80$  & $80$ & $0$    & $-20$\\\hline
C & $110$ & $110$ & $-10$  & $0$\\\hline
\end{tabular}
\end{center}
  \caption{Pivotal mechanism} \label{tab:si1}
\end{table}

\begin{table}[t]%[htbp]
\begin{center}
\begin{tabular}{|c|c|c|c|c|}
\hline
player & type & submitted type & tax & $u_i$ \\\hline
A & $110$  & $110$ & $0$ & $10$\\\hline
B & $80$  & $80$ & $0$    & $-20$\\\hline
C & $110$ & $300$ & $-10$  & $0$\\\hline
\end{tabular}
\end{center}
  \caption{Sequential pivotal mechanism}\label{tab:si2}
\end{table}

However, as Table \ref{tab:si2} shows, budget balance does not need to
be achieved.  The following result shows that an order can always be
found that yields budget balancedness.

\begin{theorem} \label{thm:zero}
  Let ${\cal D}$ be a public project problem with the sequential
  pivotal mechanism. For all $c > 0, n \geq 2$ and $\theta \in \Theta$
  there exists a permutation of players such that when each player $i$
  follows strategy $s_i(\cdot)$ of Theorem \ref{thm:dom1}, budget
  balance is achieved.
\end{theorem}

\Proof (Sketch).
Recall that in the pivotal
mechanism, given the sequence of types $\theta$, a player $i$ is
called \oldbfe{pivotal} if $t_i(\theta) \neq 0$.
First we show that not all players can be pivotal. Then we show that
the desired permutation is the one in which the last player is not pivotal.
\HB
\III

%We relegate the proof to the Appendix.  
For instance, in Example \ref{exa:tables} when the order is A, C, B or
C, A, B, the decision is taken with no taxes incurred, i.e., budget
balance is then achieved.

% In general, if $i$ is the first player for which
% $\sum_{j=1}^{i}\theta_j \geq c$, then he will submit $c$ according to
% strategy $s_i(\cdot)$. This reduces the taxes of all players except
% his own to 0.
% %(\emph{Case 2} in the proof of Theorem \ref{thm:dom1}$(ii)$).  
% Player's $i$ tax may or may not become 0. If
% he is not the last player, then all players $i+1, \LL, n$ following
% him also submit $c$, which ensures that all taxes \emph{including} the
% one of player $i$ become 0.

In Theorem \ref{thm:dom1}$(ii)$ we seem to be maximizing the social
welfare.  However, this is not the case
because we assume there that each player submits a
type that does not alter the decision to be taken.
In fact, strategy $s_i(\cdot)$ of Theorem
\ref{thm:dom1} is not socially optimal.

The following theorem provides a socially optimal strategy that in
some circumstances yields a higher social welfare than the above
strategy.

\begin{theorem} \label{thm:social}
  Let ${\cal D}$ be a public project problem. Let
\[
s_i(\theta_1, \LL, \theta_i) :=
\left\{
        \begin{array}{l@{\extracolsep{3mm}}l}
        \theta_i    & \mathrm{if}\  \sum_{j=1}^{i}\theta_j < c\mathrm{\ and \ } i<n\mbox{,} \\
       0  & \mathrm{if}\  \sum_{j=1}^{i}\theta_j < c\mathrm{\ and \ } i=n\mbox{,} \\
       0  & \mathrm{if}\  \sum_{j=1}^{i}\theta_j = c, \ \theta_i > \frac{c}{n}
\mathrm{\ and \ } i=n\mbox{,} \\
       c       & \mathrm{otherwise}
        \end{array}
        \right.
\]
be a strategy for player $i$. Then

\begin{enumerate} \smallromani

\item $s_i(\cdot)$ is socially
  optimal for player $i$ in the sequential pivotal mechanism,

\item 
for all $\theta \in \Theta$ and vectors $s'(\cdot)$ of optimal players' strategies,
\[
SW(\theta, s(\cdot)) \geq SW(\theta, s'(\cdot)),
\]
where $s(\cdot)$ is the vector of strategies $s_i(\cdot)$.
\end{enumerate}
\end{theorem}

%We relegate the proof to the Appendix.  
The remarkable thing about the
above strategy $s_i(\cdot)$ is that when $\sum_{j=1}^{n}\theta_j = c$
and $\theta_n > \frac{c}{n}$, player $n$ submits type 0, as a result
of which the project does not take place.  To illustrate this
situation reconsider Example \ref{exa:tables}. When the players submit
their types sequentially in order A, B, C following the above
strategy $s_i(\cdot)$, then player C submits 0. The resulting social
welfare is 0 as opposed to $-10$ which results when all players follow
strategy $s_i(\cdot)$ of Theorem \ref{thm:dom1} (see Table
\ref{tab:si2}). This also shows that the latter strategy is not
socially optimal.

However, in general strategy $s_i(\cdot)$ of Theorem
\ref{thm:social} does not need to ensure budget balance.

\begin{example} \label{exa:nobalance}
Suppose that there are three
players, A, B, and C, whose true types are 60, 70, and
250, respectively, while $c$ remains 300.  When the players submit their
types following strategy $s_i(\cdot)$ of Theorem \ref{thm:social},
we get the situation summarized in  Table \ref{tab:nobalance}.

\begin{table}[htbp]
\begin{center}
\begin{tabular}{|c|c|c|c|c|}
\hline
player & type & submitted type & tax & $u_i$ \\\hline
A & $60$  & $60$  &  $0$ & $-40$\\\hline
B & $70$  & $70$  &  $0$ & $-30$\\\hline
C & $250$ & $300$ & $-70$  &  $80$\\\hline
\end{tabular}
\end{center}
  \caption{Sequential pivotal mechanism}\label{tab:nobalance}
\end{table}

Here the same decision is taken as when each player is truthful and
in both situations the deficit is $-70$.
\HB
\end{example}

On other other hand, part $(ii)$ shows that when we limit ourselves to
optimal strategies and each player follows the introduced strategy
$s_i(\cdot)$, then a maximal social welfare results.  The restriction
to the vectors of optimal strategies is necessary.  Indeed, Table
\ref{tab:nobalance} of Example \ref{exa:nobalance} shows that when the
order is A, B, C and each player follows the strategy $s_i(\cdot)$ of
Theorem \ref{thm:social}, then the resulting social welfare is $380 -
300 - 70 = 10$. However, when player B submits 300, then player $C$
pays no tax and the resulting social welfare is higher, namely $380 -
300 = 80$.

% \begin{example}
% \mbox{}
% \begin{table}[htbp]

% Consider the two situations depicted in the two tables below.
% \begin{center}
% \begin{tabular}{|c|c|c|c|c|}
% \hline
% player & type & submitted type & tax & $u_i$ \\\hline
% A & $110$  & $110$ & $-10$ & $0$\\\hline
% B & $80$  & $80$ & $0$    & $-20$\\\hline
% C & $110$ & $110$ & $-10$  & $0$\\\hline
% \end{tabular}\vspace{0.25cm}
% \end{center}
%   \caption{Project takes place}
%   \label{tab:a}
% \end{table}

% \begin{table}[htbp]
% \begin{center}
% \begin{tabular}{|c|c|c|c|c|}
% \hline
% player & type & submitted type & tax & $u_i$ \\\hline
% A & $110$  & $110$ & $0$ & $0$\\\hline
% B & $80$  & $80$ & $0$   & $0$\\\hline
% C & $110$ & $0$ & $0$  & $0$\\\hline
% \end{tabular}\vspace{0.25cm}
% \end{center}
%   \caption{Project does not take place}
%   \label{tab:b}
% \end{table}

% So by submitting the type 0 instead of the true type, 110, player C increased the social welfare
% from $-20$ to 0.
% \HB
% \end{example}

\section{Comments on a Nash implementation}
\label{sec:nash}

The sequential mechanisms here considered circumvent the limitations
of the customary, simultaneous, Groves mechanisms. This and the fact
that we maximize social welfare by using strategies that deviate from
truth-telling requires some clarification.  First of all, we can
explain these sequential mechanisms by turning them into simultaneous
ones as follows.

We assume that each player $i$ receives a type $\theta_i \in \Theta_i$
and subsequently submits a function $r_i: \Theta_1 \times \LL \times
\Theta_{i-1} \myra \Theta_i$ instead of a type $\theta'_i \in
\Theta_i$.  (In particular, player 1 submits a type, i.e., $r_1(\cdot)
\in \Theta_1$.)  The submissions are simultaneous. Then the behaviour of
player $i$ can be described by a strategy $s_i: \Theta_1 \times \LL \times
\Theta_i \myra \Theta_i$ which when applied to the received type $\theta_i$ 
yields the function $s_i(\cdot, \theta_i): \Theta_1 \times
\LL \times \Theta_{i-1} \myra \Theta_i$ that player $i$ submits.  Then
$\theta$ and the vector $s(\cdot) := (s_1(\cdot), \LL, s_n(\cdot))$ of
strategies that the players follow yield an element $[s(\cdot),
\theta]$ of $\Theta$, where, recall, $[s(\cdot), \theta]_1 :=
s_{1}(\theta_1)$ and $[s(\cdot), \theta]_{i+1} := s_{i+1}([s(\cdot),
\theta]_1, \LL, [s(\cdot), \theta]_i, \theta_{i+1})$.

Given a decision problem ${\cal D} := (D, \Theta_1, \LL, \Theta_n, v_1, \LL, v_n, f)$
and two strategies $s_i(\cdot)$ and $s'_i(\cdot)$ of player $i$
in the sequential version of ${\cal D}$, we write

\begin{tabbing}
\quad $s_i(\cdot) \geq_d s'_i(\cdot)$ iff \= for all $\theta \in \Theta$ \\
\> $v_i(f(s_i(\theta_1,\LL, \theta_i), \theta_{-i}), \theta_i) \geq
v_i(f(s'_i(\theta_1,\LL, \theta_i), \theta_{-i}), \theta_i)$.
\end{tabbing}

\NI
We write $s_i(\cdot) >_d s'_i(\cdot)$ if additionally one of these inequalities is strict,
and we write $s_i(\cdot) =_d s'_i(\cdot)$ if all these inequalities are equalities.

Note that $s_i(\cdot) \geq_d s'_i(\cdot)$ for all strategies $s'_i(\cdot)$ of player $i$ iff
strategy $s_i(\cdot)$ of player $i$ is optimal in the sequential version of ${\cal D}$.

Next, we define for all $i \in \{1, \LL, n\}$ a preference relation $\succeq_i$ on
the vectors of players' strategies by writing
\begin{tabbing}
\quad $s(\cdot) \succeq_i s'(\cdot)$ iff \= $s_i(\cdot) >_d s'_i(\cdot)$ or \\
                                         \> ($s_i(\cdot) =_d s'_i(\cdot)$ and \\
                                         \> \ for all $\theta \in \Theta$, $v_i(f([s(\cdot), \theta]), \theta_i) \geq v_i(f([s'(\cdot), \theta]), \theta_i)$).
\end{tabbing}

We now say that a joint strategy $s(\cdot)$ is a \oldbfe{Nash
  equilibrium} in the sequential version of ${\cal D}$ if for all $i \in \{1, \LL, n\}$ and all
strategies $s'_i(\cdot)$ of player $i$ we have
\[
(s_i(\cdot), s_{-i}(\cdot)) \succeq_i (s'_i(\cdot), s_{-i}(\cdot)).
\]

% Note that not all vectors of optimal strategies form a Nash
% equilibrium in the sequential version of the pivotal mechanism for the
% public project problem.
% Indeed, consider the following strategy

% \[
% s'_i(\theta_1, \LL, \theta_i) :=
%  \left\{
%         \begin{array}{l@{\extracolsep{1mm}}l}
%         c           & \mathrm{if}\  i > 1 \ \mathrm{and} \ \theta_{i-1} = c \\
%         \theta_i    & \mathrm{otherwise}
%         \end{array}
%         \right.
% \]

% By Lemma \ref{lem:vcg1} each $s'_i(\cdot)$ strategy is optimal.
% However, $(s'_1(\cdot), \LL, s'_n(\cdot))$ is not a Nash equilibrium.  The
% reason is that if a player $i \in \{1, \LL, n-1\}$ deviates to
% strategy $s_i(\cdot)$ of Theorem \ref{thm:dom1}, then he is strictly
% better off for $\theta \in \Theta$ such that $\sum_{j=1}^{i-1}
% \theta_{j} < \frac{n-1}{n} c$, $\theta_i = c -
% \sum_{j=1}^{i-1}\theta_{j}$, and $\theta_k = 0$ for $k \in \{i+1, \LL,
% n\}$.  Indeed, the final utility of player $i$ then increases from
% $\theta_i - \frac{c}{n} + (\sum_{j=1}^{i-1} \theta_{j} - \frac{n-1}{n}
% c) = 0$ to $\theta_i - \frac{c}{n}$ because in the case of deviation
% the tax of player $i$ becomes 0.

% On the other hand, we have the following result that clarifies the
The following result clarifies the status of the strategies introduced
in Theorems \ref{thm:dom1} and \ref{thm:social}.

\begin{theorem} \label{thm:nash}
  Let ${\cal D}$ be a public project problem. 
  \begin{enumerate} \smallromani
  \item Each of the vectors
  $s(\cdot)$ of strategies defined in Theorems
  \ref{thm:dom1} and \ref{thm:social}, respectively, is a Nash
  equilibrium in the corresponding sequential version of the pivotal
  mechanism.
  
\item The vector $s(\cdot)$ of Theorem \ref{thm:dom1} is Pareto
  optimal in the universe of optimal strategies, in the sense that for
  all $\theta \in \Theta$ the resulting social welfare $SW(\theta,
  s(\cdot))$ is maximal among all vectors of optimal players'
  strategies.
  \end{enumerate}
\end{theorem}

%The proof can be found in the Appendix.
This result shows that the improvement in terms of the
maximization of the social welfare over the Groves mechanism is
achieved by weakening the implementation in dominant strategies (see
Groves Theorem) to an implementation in Nash equilibrium
(in the universe of optimal strategies).

The above definition of the $\succeq_i$ relation uses the $>_d$
relation to ensure that in the definition of Nash equilibrium the
deviations to non-optimal strategies are trivially discarded.  This
ruling out of non-optimal strategies is necessary.  Indeed, when
$\theta_{i} > \frac{c}{n}$, with $i < n$, and $\sum_{j =
  1}^{n}\theta_j < c$, then player's $i$ final utility increases from
0 to $\theta_{i} - \frac{c}{n}$ when he deviates from any of the two
strategies considered in Theorem \ref{thm:nash} to the strategy
\[
s_i(\theta_1,\LL, \theta_i) :=
 \left\{
        \begin{array}{l@{\extracolsep{3mm}}l}
        0    & \mathrm{if}\  \theta_i \leq  \frac{c}{n} \\
        c       & \mathrm{otherwise}.
        \end{array}
        \right.
\]

Recall now that the well-known revelation principle (see, e.g.,
\cite{Mye91}) states that every mechanism can be realized as a
(simultaneous) direct mechanism in which truth-telling is the optimal
strategy.  We now show that using any Nash equilibrium $(s_1(\cdot),
\LL, s_n(\cdot))$ of Theorem \ref{thm:nash} we can construct a
revelation-type simultaneous mechanism in which the vector
$(\pi_1(\cdot), \LL, \pi_n(\cdot))$ of the projection functions forms
a Nash equilibrium.  (Recall that the $\pi_i(\cdot)$ function
corresponds in the sequential setting to truth-telling by player $i$.)
This mechanism is constructed using the following preference relations
$\succeq^{*}_i$ on the vectors of players' strategies:

\begin{tabbing}
\quad $s'(\cdot) \succeq^{*}_i s''(\cdot)$ iff \\
\quad $(s_1(\cdot) \circ s'_1(\cdot), \LL, s_n(\cdot) \circ s'_n(\cdot)) \succeq_i (s_1(\cdot) \circ s''_1(\cdot), \LL, s_n(\cdot) \circ s''_n(\cdot))$,
\end{tabbing}

\NI
where strategy $s_i(\cdot) \circ s'_i(\cdot)$ of player $i$ is defined by
\[
(s_i(\cdot) \circ s'_i(\cdot))(\theta_1, \LL, \theta_i) := s_i (\theta_1, \LL, \theta_{i-1}, s'_i(\theta_1, \LL, \theta_i)).
\]

\begin{theorem} \label{thm:nash1}
  Let ${\cal D}$ be a public project problem. 
The vector $(\pi_1(\cdot), \LL, \pi_n(\cdot))$ of projection strategies
is a Nash equilibrium in the corresponding sequential version
of the pivotal mechanism, where we use the preference relations
$\succeq^{*}_1, \LL, \succeq^{*}_n$.
\end{theorem}

\Proof
Note that for all $j \in \{1, \LL, n\}$,
$s_j(\cdot) \circ \pi_j(\cdot) = s_j (\cdot)$. Then
\[
\mbox{$(\pi_i(\cdot), \pi_{-i}(\cdot)) \succeq^{*}_i (s'_i(\cdot), \pi_{-i}(\cdot))$ iff
$(s_i(\cdot), s_{-i}(\cdot)) \succeq_i (s_i(\cdot) \circ s'_i(\cdot), s_{-i}(\cdot))$},
\]
so the result holds by Theorem \ref{thm:nash}$(i)$.
\HB

\section{Concluding remarks}
\label{sec:conc}

As already mentioned, no budget balanced Groves mechanisms exist for the
public project. We have investigated here to what extent the
unavoidable deficit can be reduced when players move sequentially.  By
focusing on socially optimal strategies we have
incorporated into our analysis altruistic behaviour of the players.

% The following quote from \cite[ page 109]{Bow04} can be of interest
% here (both emphases in the text): ``\emph{Other-regarding} preferences
% include spite, altruism, and caring about the relationship among the
% outcomes for oneself and others. [$\dots$] \emph{The key aspect of
%   other-regarding preferences is that one's evaluation of a state
%   depends on how it is experienced by others.}''  Bowles also provides
% the following elegant quote from Dalai Lama: ``The intelligent way to
% be selfish is to work for the welfare of others''.

The results here established hold for the sequential pivotal mechanism.
Some of them, but not all, can be generalized to sequential Groves
mechanisms.  More specifically, the strategies introduced in 
Theorems \ref{thm:dom1} and \ref{thm:social} are also optimal
in arbitrary sequential Groves mechanisms. The reason is the
following observation.

\begin{note}
  Fix an initial decision problem and consider two Groves mechanisms
  (with tax functions) $t$ and $t'$.  A strategy of player $i$ is optimal in the sequential
  version of $t$ iff it is optimal in the sequential version of $t'$.
\end{note}
% \Proof 
% Given the vector $\theta_{-i}$ of announced types of the other
% players, the final utilities of player $i$ in two Groves mechanisms
% differ by a function of $\theta_{-i}$.  This implies the claim by
% definition of optimality.  
% \HB 
% \II

How to generalize the remaining claims of Theorems \ref{thm:dom1}
and \ref{thm:social} to other sequential Groves mechanisms remains an
interesting open problem.

%\bibliography{/ufs/apt/bib/e,/ufs/apt/bib/apt}
%\bibliographystyle{plain}

\section*{Appendix}

We provide here the proofs of Lemma \ref{lem:vcg1} and Theorems \ref{thm:dom1}, \ref{thm:zero}, 
\ref{thm:social} and \ref{thm:nash}.
\III

\NI
\textbf{Proof of Lemma \ref{lem:vcg1}}.

We have for all $\theta \in \Theta$ and
$\theta'_i\in\Theta_i$
\begin{align*}
&\phantom{= \ \:}  u_i((f,t)(s_i(\theta_1,\LL, \theta_i), \theta_{-i}), \theta_i)  \\
&= \sum_{j=1}^{n} v_j(f(s_i(\theta_1,\LL,\theta_i), \theta_{-i}), \theta_j) + h_i(\theta_{-i}) \\
&= \sum_{j=1}^{n} v_j(f(\theta_i, \theta_{-i}), \theta_j)  + h_i(\theta_{-i}) \\
&= u_i((f,t)(\theta_i, \theta_{-i}), \theta_i) \\
&\geq u_i((f,t)(\theta'_i, \theta_{-i}), \theta_i),
\end{align*}
where the first equality holds by the definition of Groves
mechanisms, the second equality holds by our assumption, and the
inequality follows by Groves Theorem.
\HB 

Below we frequently use the following observation.

\begin{note} \label{not:t}
Let ${\cal D}$ be a public project problem with
the sequential pivotal mechanism.
Then
\[
t_i(\theta'_i, \theta_{-i}) =
 \left\{
        \begin{array}{l@{\extracolsep{3mm}}l}
        \min(0, \frac{n-1}{n} c - \sum_{k \neq i}\theta_k)
 & \mathrm{if}\  \sum_{k \neq i}\theta_k + \theta'_i < c \\
        \min(0, \sum_{k \neq i}\theta_k - \frac{n-1}{n} c)
& \mathrm{otherwise}
        \end{array}
        \right.
\]

% \begin{enumerate} \smallromani

% \item Suppose $f(\theta'_i, \theta_{-i}) = 1$. Then
% \[
% t_i(\theta'_i, \theta_{-i}) =
%  \left\{
%         \begin{array}{l@{\extracolsep{3mm}}l}
%        0  & \mathrm{if}\  \sum_{j \neq i} \theta_j \geq  \frac{n-1}{n} c \\
%        \sum_{j \neq i}\theta_j - \frac{n-1}{n} c       & \mathrm{otherwise.}
%         \end{array}
%         \right.
% \]

% \item Suppose $f(\theta'_i, \theta_{-i}) = 0$. Then
% \[
% t_i(\theta'_i, \theta_{-i}) =
%  \left\{
%         \begin{array}{l@{\extracolsep{3mm}}l}
%        0  & \mathrm{if}\  \frac{n-1}{n} c \geq \sum_{j \neq i} \theta_j \\
%        \frac{n-1}{n} c - \sum_{j \neq i}\theta_j        & \mathrm{otherwise.}
%         \end{array}
%         \right.
% \]
% \HB
% \end{enumerate}
\HB
\end{note}

\NI
\textbf{Proof of Theorem \ref{thm:dom1}}.

\NI
$(i)$ By Lemma \ref{lem:vcg1} it suffices to show
that
$f(s_i(\theta_1,\LL,\theta_i), \theta_{-i})=f(\theta_i, \theta_{-i})$. For this we consider three cases.
\II

\NI
\emph{Case 1} $s_i(\theta_1,\LL,\theta_i)=\theta_{i}$.

Then $f(s_i(\theta_1,\LL,\theta_i), \theta_{-i})=f(\theta_i, \theta_{-i})$.
\II

\NI
\emph{Case 2} $s_i(\theta_1,\LL,\theta_i) = 0$.

By definition of $s_i(\cdot)$, we have $i=n$
and $c>\sum_{j=1}^{n}\theta_j \geq s_i(\theta_1,\LL,\theta_i)+\sum_{i\not=j} \theta_j$
and therefore $f(s_i(\theta_1,\LL,\theta_i), \theta_{-i})=f(\theta_i, \theta_{-i})$,
as both sides equal 0.
\II

\NI
\emph{Case 3} $s_i(\theta_1,\LL,\theta_i) = c$.

By definition of $s_i(\cdot)$, we have both
$\sum_{j=1}^{n}\theta_j \geq \sum_{j=1}^{i}\theta_j \geq c$ and
$s_i(\theta_1,\LL,\theta_i)+\sum_{i\not=j} \theta_j \geq c$, so $f(s_i(\theta_1,\LL,\theta_i), \theta_{-i})=f(\theta_i, \theta_{-i})$,
as both sides equal 1.
\II

\NI
$(ii)$ By Note \ref{not:t} we have for all $\theta \in \Theta$, $j \neq i$ and $\theta'_i \in \Theta_i$
\[
t_j(\theta'_i, \theta_{-i}) =
 \left\{
        \begin{array}{l@{\extracolsep{3mm}}l}
        \min(0, \frac{n-1}{n} c - (\sum_{k \neq i,j}\theta_k + \theta'_i))
 & \mathrm{if}\  \sum_{k \neq i}\theta_k + \theta'_i < c \\
        \min(0, \sum_{k \neq i,j}\theta_k + \theta'_i - \frac{n-1}{n} c)
& \mathrm{otherwise}
        \end{array}
        \right.
\]

Assume that $s_i(\theta_1,\LL,\theta_i) \neq \theta_i$ and
$f(\theta'_i, \theta_{-i}) = f(\theta_i, \theta_{-i})$.
Two cases arise.
\II

\NI
\emph{Case 1} $s_i(\theta_1,\LL,\theta_i) = 0$.

By definition of $s_i(\cdot)$ we have
$\sum_{k=1}^{n}\theta_k < c$. Then $\sum_{k \neq i}\theta_k + s_i(\theta_1,\LL,\theta_i) < c$.
Also
$\sum_{k \neq i}\theta_k + \theta'_i < c$ since
$f(\theta'_i, \theta_{-i}) = f(\theta_i, \theta_{-i}) = 0$.
Hence
\[
t_j(s_i(\theta_1,\LL,\theta_i), \theta_{-i}) = \min(0, \frac{n-1}{n} c - (\sum_{k \neq i,j}\theta_k + 0)) \geq
\]
\[
\min(0, \frac{n-1}{n} c - (\sum_{k \neq i,j}\theta_k + \theta'_i)) = t_j(\theta'_i, \theta_{-i}).
\]

\NI
\emph{Case 2} $s_i(\theta_1,\LL,\theta_i) = c$.

Then $\sum_{k \neq i}\theta_k + s_i(\theta_1,\LL,\theta_i) \geq c$.
Hence
\[
t_j(s_i(\theta_1,\LL,\theta_i), \theta_{-i}) = \min(0, \sum_{k \neq i,j}\theta_k + c - \frac{n-1}{n} c) = 0
\geq t_j(\theta'_i, \theta_{-i}),
\]
where the last inequality holds since the pivotal mechanism is pay only.

Now, in $(i)$ we showed that $f(s_i(\theta_1,\LL,\theta_i),
\theta_{-i})=f(\theta_i, \theta_{-i})$ and by assumption
$f(\theta'_i, \theta_{-i}) = f(\theta_i, \theta_{-i})$. 

% When $\theta_i$ is the received type of player $i$ 
% and his announced type is $\theta'_i$, then
% \begin{equation}
%   \label{equ:u}
% u_i((f,t)(\theta'_i,\theta_{-i}), \theta_i) = 
% v_i(f(\theta'_i, \theta_{-i}), \theta_i) + t_i(\theta'_i, \theta_{-i}),
% \end{equation}
% where $\theta_{-i}$ are the types announced by the other players.

Hence by the definition of $u_j$
\[
u_j((f,t)(s_i(\theta_1,\LL, \theta_i), \theta_{-i}), \theta_j) 
= v_j(f(\theta_i, \theta_{-i}), \theta_i) + t_j(s_i(\theta_1,\LL,\theta_i), \theta_{-i})
\]
and
\[
u_j((f,t)(\theta'_i, \theta_{-i}), \theta_j) 
= v_j(f(\theta_i, \theta_{-i}), \theta_i) + t_j(\theta'_i, \theta_{-i}),
\]
which yields the claim.
% \[
% u_j((f,t)(s_i(\theta_1,\LL, \theta_i), \theta_{-i}), \theta_j) \geq
% u_j((f,t)(\theta'_i, \theta_{-i}), \theta_j).
% \]
\HB
\II

\NI
\textbf{Proof of Theorem \ref{thm:zero}}.
\II

% Recall that in the pivotal mechanism, given the sequence of types $\theta$, a player $i$ is called \oldbfe{pivotal}
% if $t_i(\theta) \neq 0$, i.e., $f(\theta) \not\in \textrm{argmax}_{d\in D} \sum_{j \neq i} v_j(d, \theta_j)$.
We distinguish two cases.
\II

\NI
\emph{Case 1} $f(\theta) = 1$.

Then a player $i$ is pivotal iff
$\sum_{j \neq i} v_j(1, \theta_j) < \sum_{j \neq i} v_j(0, \theta_j)$,
i.e., iff $\sum_{j \neq i} \theta_j < \frac{n-1}{n} c$.

This means that not all players can be pivotal, since then we would have
\[
(n-1) \sum_{k = 1}^{n} \theta_k = \sum_{i = 1}^{n} \sum_{j \neq i} \theta_j < n \frac{n-1}{n} c = (n-1) c,
\]
which cannot be the case as we have assumed that $f(\theta) = 1$.

Choose then a permutation of the players in which the last player is
not pivotal and suppose that each player $i$ submits type $\theta'_i$
following the strategy $s_i(\cdot)$ of Theorem \ref{thm:dom1}.
Then the last player submits $\theta'_n := c$.

This will make  all players non-pivotal for the sequence of types
$\theta'_1, \LL, \theta'_n$. Indeed, for $i \neq n$ we have then
$\sum_{j \neq i} \theta'_j \geq c > \frac{n-1}{n} c$, and for $i = n$ we have
$\sum_{j \neq i} \theta'_j \geq \sum_{j \neq i} \theta_j \geq \frac{n-1}{n} c$,
where the second inequality holds by the definition of the strategies $s_j(\cdot)$ and by the fact that
$n$ is not pivotal. Therefore all taxes become 0.
\II

\NI
\emph{Case 2}  $f(\theta) = 0$.

Here, a player $i$ is pivotal iff
$\sum_{j \neq i} \theta_j > \frac{n-1}{n} c$.
By an analogous argument as in Case 1, not all players can be pivotal.
% Then a player $i$ is pivotal iff
% $\sum_{j \neq i} v_j(0, \theta_j) < \sum_{j \neq i} v_j(1, \theta_j)$,
% i.e., iff $\sum_{j \neq i} \theta_j > \frac{n-1}{n} c$.
%
% This means that not all players can be pivotal, since then we would have
% \[
% (n-1) \sum_{k = 1}^{n} \theta_k = \sum_{i = 1}^{n} \sum_{j \neq i} \theta_j > n \frac{n-1}{n} c = (n-1) c,
% \]
% which cannot be the case as we assumed that $f(\theta) = 0$.
%
Choose then a permutation of the players in which the last player is
not pivotal and suppose that each player $i$ submits his type $\theta'_i$
following the strategy $s_i(\cdot)$ of Theorem \ref{thm:dom1}.
Then the last player submits $\theta'_n := 0$.

This makes all players non-pivotal for the sequence of types
$\theta'_1, \LL, \theta'_n$. Indeed,
by definition of the strategies $s_j(\cdot)$ and the fact that $n$ is not pivotal
we have then
$\sum_{j \neq i} \theta'_j \leq \sum_{j \neq i} \theta_j - \theta_n \leq \sum_{j \neq n} \theta_j \leq
\frac{n-1}{n} c$ for $i \neq n$, and
$\sum_{j \neq n} \theta'_j \leq \sum_{j \neq n} \theta_j \leq \frac{n-1}{n} c$.
Therefore all taxes become 0.
\HB
\II

The other proofs rely on the following lemma that clarifies the status of optimal strategies.

\begin{lemma} \label{lem:compat}
  Let ${\cal D}$ be a public project problem. and let $s'_i(\cdot)$ be
  an optimal strategy for player $i$ in the corresponding sequential
  pivotal mechanism.

\begin{enumerate} \smallromani
\item Suppose  $\sum_{j=1}^{i}\theta_j < c$ and $i < n$. Then $s'_i(\theta_1, \LL, \theta_i) = \theta_i$.

\item Suppose $\sum_{j=1}^{i}\theta_j < c$ and $i = n$. Then $\sum_{j=1}^{n-1}\theta_j + s'_i(\theta_1, \LL, \theta_n)
 < c$.

\item Suppose $\sum_{j=1}^{i}\theta_j = c$ and $i < n$.
Then $s'_i(\theta_1, \LL, \theta_i) \geq \theta_i$.

\item Suppose $\sum_{j=1}^{i}\theta_j > c$. Then $\sum_{j=1}^{i-1}\theta_j +
s'_i(\theta_1, \LL, \theta_i) \geq c$.

\end{enumerate}

\end{lemma}

\Proof Denote $s'_i(\theta_1, \LL, \theta_i)$ by $\theta'_i$.

\NI
$(i)$
We proceed by contradiction. Assume that $\theta'_i \neq \theta_i$. We consider two cases.
\II

\NI
\emph{Case 1} $\theta'_i > \theta_i$.

Choose $\theta_{i+1} \in \Theta_{i+1}, \LL, \theta_n \in \Theta_n$ so that $\sum_{j = 1}^{n} \theta_j < c$ and
$\sum_{j \neq i}\theta_j + \theta'_i \geq c$. Then $f(\theta) = 0$ and $f(\theta'_i, \theta_{-i}) = 1$.
\II

\NI
\emph{Subcase 1} $\sum_{j \neq i}\theta_j < \frac{n-1}{n} c$.

Then $u_i((f,t)(\theta), \theta_i) = t_i(\theta) = 0$.
Moreover, 
$
u_i((f,t)(\theta'_i, \theta_{-i}), \theta_i) = (\theta_i - \frac{c}{n}) + t_i(\theta'_i, \theta_{-i}) =
(\theta_i - \frac{c}{n}) + (\sum_{j \neq i}\theta_j - \frac{n-1}{n} c) =
\sum_{j = 1}^{n} \theta_j - c < 0$.
\II

\NI
\emph{Subcase 2} $\sum_{j \neq i}\theta_j \geq \frac{n-1}{n} c$.

Then $u_i((f,t)(\theta), \theta_i) = t_i(\theta) = \frac{n-1}{n} c - \sum_{j \neq
  i}\theta_j$.
Moreover, 
\[
u_i((f,t)(\theta'_i, \theta_{-i}), \theta_i) = (\theta_i -
\frac{c}{n}) + t_i(\theta'_i, \theta_{-i}) = \theta_i - \frac{c}{n} <
\frac{n-1}{n} c - \sum_{j \neq i}\theta_j,
\]
where the last inequality
holds since $\sum_{j = 1}^{n} \theta_j - c < 0$.
\II

\NI
\emph{Case 2} $\theta'_i < \theta_i$.

Choose $\theta_{i+1} \in \Theta_{i+1}, \LL, \theta_n \in \Theta_n$ so that
$\sum_{j = 1}^{n} \theta_j > c$ and
$\sum_{j \neq i}\theta_j + \theta'_i < c$. Then $f(\theta) = 1$ and $f(\theta'_i, \theta_{-i}) = 0$.
\II

\NI
\emph{Subcase 1} $\sum_{j \neq i}\theta_j < \frac{n-1}{n} c$.

Then $u_i((f,t)(\theta), \theta_i) = (\theta_i - \frac{c}{n}) + t_i(\theta) =
(\theta_i - \frac{c}{n}) + (\sum_{j \neq i}\theta_j - \frac{n-1}{n} c) =
\sum_{j = 1}^{n} \theta_j - c > 0$.
Moreover $u_i((f,t)(\theta'_i, \theta_{-i}), \theta_i) = t_i(\theta'_i, \theta_{-i}) = 0$.
\II

\NI
\emph{Subcase 2} $\sum_{j \neq i}\theta_j \geq \frac{n-1}{n} c$.

Then $u_i((f,t)(\theta), \theta_i) = (\theta_i - \frac{c}{n}) + t_i(\theta) = \theta_i
- \frac{c}{n}$.
Moreover $u_i((f,t)(\theta'_i, \theta_{-i}), \theta_i) =
t_i(\theta'_i, \theta_{-i}) = \frac{n-1}{n} c - \sum_{j \neq
  i}\theta_j < \theta_i - \frac{c}{n}$, where the last inequality
holds since $c - \sum_{j = 1}^{n} \theta_j < 0$.
\II

In both cases we have showed that $u_i((f,t)(\theta'_i, \theta_{-i}), \theta_i) < u_i((f,t)(\theta), \theta_i)$,
i.e., strategy $s'_i(\cdot)$ is not optimal.
\III

\NI
$(ii)$ We proceed by contradiction. Assume that $\sum_{j=1}^{n-1}\theta_j + \theta'_n \geq c$.
Then $f(\theta) = 0$ and $f(\theta'_n, \theta_{-n}) = 1$ and as in Case 1 of $(i)$ we conclude that
$u_n((f,t)(\theta'_n, \theta_{-n}), \theta_n) < u_n((f,t)(\theta), \theta_n)$. Hence
strategy $s'_n(\cdot)$ is not optimal.
\III

\NI
$(iii)$
We proceed by contradiction. Assume that $\theta'_i < \theta_i$.

\NI
\emph{Case 1} $\theta_i > \frac{c}{n}$.
\II

\NI
\emph{Subcase 1} $\frac{c}{n} \leq \theta'_i$.

Choose $\theta_{i+1} \in \Theta_{i+1}, \LL, \theta_n
\in \Theta_n$ so that $\sum_{j \neq i}\theta_j + \theta'_i < c$
and $\sum_{j =1}^{n} \theta_j > c$.
Then $f(\theta) = 1$ and
$f(\theta'_i, \theta_{-i}) = 0$.
Besides, $\sum_{j \neq i}\theta_j < \frac{n-1}{n} c$ since  $\frac{c}{n} \leq \theta'_i$.

Hence $u_i((f,t)(\theta), \theta_i) = (\theta_i - \frac{c}{n}) +
t_i(\theta) = (\theta_i - \frac{c}{n}) +  (\sum_{j \neq i}\theta_j - \frac{n-1}{n} c) =
\sum_{j =1}^{n} \theta_j -c > 0$.
On the other hand
$u_i((f,t)(\theta'_i, \theta_{-i}), \theta_i) =
t_i(\theta'_i, \theta_{-i}) \leq 0$, since
the pivotal mechanism is pay only.
Hence strategy $s'_i(\cdot)$ is not optimal.
\II

\NI
\emph{Subcase 2} $\frac{c}{n} > \theta'_i$.

Choose $\theta_{i+1} \in \Theta_{i+1}, \LL, \theta_n
\in \Theta_n$ so that $\sum_{j \neq i}\theta_j + \theta'_i < c$ and
$\sum_{j \neq i} \theta_j > \frac{n-1}{n} c$.  Then $f(\theta) = 1$ and
$f(\theta'_i, \theta_{-i}) = 0$.

Hence $u_i((f,t)(\theta), \theta_i) = (\theta_i - \frac{c}{n}) +
t_i(\theta) = \theta_i - \frac{c}{n} >0$, while 
\[
u_i((f,t)(\theta'_i, \theta_{-i}), \theta_i) = t_i(\theta'_i, \theta_{-i}) \leq 0,
\] 
since the pivotal mechanism is pay only.
Therefore strategy $s'_i(\cdot)$ is not optimal.
\II

\NI
\emph{Case 2} $\theta_i \leq \frac{c}{n}$.

Recall that we assumed that $\theta'_i < \theta_i$.
Choose $\theta_{i+1} \in \Theta_{i+1}, \LL, \theta_n \in \Theta_n$ so
that $\sum_{j = 1}^{n} \theta_j > c$ and $\sum_{j \neq
  i}\theta_j + \theta'_i < c$.  Then $f(\theta) = 1$ and $f(\theta'_i,
\theta_{-i}) = 0$.  Hence $u_i((f,t)(\theta), \theta_i) = \theta_i -
\frac{c}{n}$, while $u_i((f,t)(\theta'_i, \theta_{-i}), \theta_i) =
t_i(\theta'_i, \theta_{-i}) = \frac{n-1}{n} c - \sum_{j \neq
  i}\theta_j$, since by the assumptions $\frac{n-1}{n} c < \sum_{j \neq i}\theta_j$.

But we have $\theta_i - \frac{c}{n} > \frac{n-1}{n} c - \sum_{j \neq
  i}\theta_j$, so $u_i((f,t)(\theta), \theta_i) > u_i((f,t)(\theta'_i,
\theta_{-i})$, i.e., strategy $s'_i(\cdot)$ is not optimal.
\III

\NI
$(iv)$ We proceed by contradiction. Assume that
$\sum_{j=1}^{i-1}\theta_j + \theta'_i < c$. Choose
$\theta_{i+1} \in \Theta_{i+1}, \LL, \theta_n \in \Theta_n$ so that
$\sum_{j = 1}^{n} \theta_j > c$ and
$\sum_{j \neq i}\theta_j + \theta'_i < c$.
Then $f(\theta) = 1$ and $f(\theta'_i, \theta_{-i}) = 0$.

Hence
$u_i((f,t)(\theta), \theta_i) = (\theta_i - \frac{c}{n}) + t_i(\theta)$ and
$u_i((f,t)(\theta'_i, \theta_{-i}), \theta_i) = t_i(\theta'_i, \theta_{-i})$.
By optimality of $s'_i(\cdot)$ we have
$u_i((f,t)(\theta'_i, \theta_{-i}), \theta_i) \geq u_i((f,t)(\theta), \theta_i)$, and hence
$t_i(\theta'_i, \theta_{-i}) \geq (\theta_i - \frac{c}{n}) + t_i(\theta)$.
\II

\NI
\emph{Case 1} $\sum_{j \neq i}\theta_j < \frac{n-1}{n} c$.

Then $t_i(\theta'_i, \theta_{-i}) = 0$ and
$t_i(\theta) = \sum_{j \neq i}\theta_j - \frac{n-1}{n} c$.
Hence $0 \geq (\theta_i - \frac{c}{n}) + (\sum_{j \neq i}\theta_j - \frac{n-1}{n} c)$, i.e.,
$\sum_{j = 1}^{n} \theta_j \leq c$.
\II

\NI
\emph{Case 2} $\sum_{j \neq i}\theta_j \geq \frac{n-1}{n} c$.

Then $t_i(\theta) = 0$ and $t_i(\theta'_i, \theta_{-i}) =
\frac{n-1}{n} c - \sum_{j \neq i}\theta_j$.
Hence $\frac{n-1}{n} c - \sum_{j \neq i}\theta_j \geq \theta_i - \frac{c}{n}$,
i.e., $\sum_{j = 1}^{n} \theta_j \leq c$, as well.
\II

So in both cases we get a contradiction, since $\sum_{j = 1}^{n} \theta_j > c$.
\HB
\II

\NI
\textbf{Proof of Theorem \ref{thm:social}}.
\II

\NI
$(i)$
To prove that strategy $s_i(\cdot)$ of player $i$
is optimal it suffices to show that
\begin{equation}
  \label{equ:eq}
\sum_{j=1}^{n} v_j(f(s_i(\theta_1,\LL,\theta_i), \theta_{-i}), \theta_j) =  \sum_{j=1}^{n} v_j(f(\theta_i, \theta_{-i}), \theta_j)
\end{equation}
and reuse the string of inequalities from the proof of Lemma \ref{lem:vcg1}.

In view of Theorem \ref{thm:dom1}$(i)$ we only need to consider the situation when
$\sum_{j=1}^{i}\theta_j = c, \ \theta_i > \frac{c}{n}$, and $i =n$.
But then $s_n(\theta_1, \LL, \theta_n) = 0$, so $f(0, \theta_{-n}) = 0$, and hence
(\ref{equ:eq}) holds for $i = n$ as both sides are equal 0.

To prove social optimality we also need to prove that for all optimal strategies $s'_i(\cdot)$ of player $i$ and all
$\theta \in \Theta$
\begin{equation}
  \label{equ:social}
 \begin{array}{l}
\sum_{j=1}^{n} u_j((f,t)(s_i(\theta_1,\LL, \theta_i), \theta_{-i}), \theta_j) \geq \\
\sum_{j=1}^{n} u_j((f,t)(s'_i(\theta_1,\LL, \theta_i), \theta_{-i}), \theta_j).
\end{array}
\end{equation}

Take some optimal strategy $s'_i(\cdot)$ of player $i$ and some
$\theta \in \Theta$.  Denote $s'_i(\theta_1, \LL, \theta_i)$ by
$\theta'_i$.  The proof follows a case analysis that reflects the
definition of strategy $s_{i}(\cdot)$.  \II

\NI
\emph{Case 1} $\sum_{j=1}^{i}\theta_j < c$ and $i < n$.

Then by Lemma \ref{lem:compat}$(i)$
$\theta'_i = \theta_i$.
Also $s_i(\theta_1,\LL, \theta_i) = \theta_i$ and then
(\ref{equ:social}) holds.
\II

\NI
\emph{Case 2}
$\sum_{j=1}^{i}\theta_j < c$ and $i = n$.

Then $s_n(\theta_1,\LL, \theta_n) = 0$ and 
\[
\sum_{j=1}^{n}
u_j((f,t)(s_n(\theta_1,\LL, \theta_n), \theta_{-n}), \theta_j) =
\sum_{j=1}^{n} t_j(0, \theta_{-n}).  
\]
Also by Lemma
\ref{lem:compat}$(ii)$ $\sum_{j=1}^{n-1}\theta_j + \theta'_n < c$, so
$f(\theta'_n, \theta_{-n}) = 0$ and 
\[
\sum_{j=1}^{n}
u_j((f,t)(\theta'_i, \theta_{-i}), \theta_j) = \sum_{j=1}^{n}
t_j(\theta'_n, \theta_{-n}).
\]

Note now that $f(0, \theta_{-n}) = f(\theta'_n, \theta_{-n}) = 0$ and
hence $t_n(0, \theta_{-n}) = t_n(\theta'_n, \theta_{-n})$, since by
Note \ref{not:t}$(ii)$ either both sides equal 0 or $\frac{n-1}{n} c
- \sum_{j \neq n}\theta_j$.  Further, also by Note \ref{not:t} $(ii)$,
for all $j \neq n$
\[
t_j(0, \theta_{-n}) = \min(0, \frac{n-1}{n} c - (\sum_{k \neq j}\theta_k + 0)) \geq
\]
\[
\min(0, \frac{n-1}{n} c - (\sum_{k \neq j}\theta_k + \theta'_n)) = t_j(\theta'_n, \theta_{-n}).
\]
Therefore we conclude that $t_j(0, \theta_{-n}) \geq t_j(\theta'_n, \theta_{-n})$.
Hence
$\sum_{j=1}^{n} t_j(0, \theta_{-n}) \geq \sum_{j=1}^{n} t_j(\theta'_n, \theta_{-n})$ and
consequently (\ref{equ:social}) holds.
\II

\NI
\emph{Case 3}
$\sum_{j=1}^{i}\theta_j = c, \ \theta_i > \frac{c}{n}$ and $i =n$.

Note that for all $d \in \{0, 1\}$ we have $\sum_{k=1}^{n} v_k(d,
\theta_j) = 0$. Also, the pivotal mechanism is pay only.  Therefore we
have
\[
\sum_{j=1}^{n} u_j((f,t)(\theta'_n, \theta_{-n}), \theta_j) = \sum_{j=1}^{n} v_j(f(\theta'_n, \theta_{-n}), \theta_j) +
\sum_{j=1}^{n} t_j(\theta'_n, \theta_{-n}) \leq 0.
\]

Further, $f(0, \theta_{-n}) = 0$ and hence $t_n(0, \theta_{-n}) = 0$.
Also for all $j \neq n$ we have $\sum_{k \neq i,j}\theta_k + 0 \leq
\sum_{k \neq i}\theta_k < \frac{n-1}{n} c$, so $t_j(0, \theta_{-n}) =
0$.

Since $s_n(\theta_1,\LL,\theta_n) = 0$, we have
\[
\sum_{j=1}^{n} u_j((f,t)(s_n(\theta_1,\LL,\theta_n), \theta_{-n}), \theta_j) = \sum_{j=1}^{n} v_j(f(s_n(\theta_1,\LL,\theta_n), \theta_{-n}), \theta_j) = 0,
\]
and consequently (\ref{equ:social}) holds.
\II

\NI
\emph{Case 4}
$\sum_{j=1}^{i}\theta_j = c, \ \theta_i \leq \frac{c}{n}$ and $i =n$.

Then $f(c, \theta_{-n}) = 1$ and hence
$t_n(c, \theta_{-n}) = 0$. Also for all $j \neq n$ we have $\sum_{k \neq i,j}\theta_k + c \geq
\frac{n-1}{n} c$, so $t_j(c, \theta_{-n}) = 0$.
Since $s_n(\theta_1,\LL,\theta_n) = c$, we have as in Case 3

$\sum_{j=1}^{n} u_j((f,t)(s_n(\theta_1,\LL,\theta_n), \theta_{-n}), \theta_j) = 0 \geq
\sum_{j=1}^{n} u_j((f,t)(\theta'_n, \theta_{-n}), \theta_j)$.
\II

\NI
\emph{Case 5}
($\sum_{j=1}^{i}\theta_j = c$ and $i < n$) or $\sum_{j=1}^{i}\theta_j > c$.

Then $s_i(\theta_1,\LL,\theta_i) = c$.
By Lemma \ref{lem:compat}$(iii)$ and $(iv)$ we have $\sum_{j \neq i}\theta_j + \theta'_i \geq c$.
Hence
$\sum_{j=1}^{n} u_j((f,t)(s_i(\theta_1,\LL, \theta_i), \theta_{-i}), \theta_j) =
\sum_{j=1}^{n} v_j(1, \theta_j) + \sum_{j=1}^{n} t_j(c, \theta_{-i})$
and
$\sum_{j=1}^{n} u_j((f,t)(\theta'_i, \theta_{-i}), \theta_j) =
\sum_{j=1}^{n} v_j(1, \theta_j) + \sum_{j=1}^{n} t_j(\theta'_i, \theta_{-i})$.

Note now that $f(c, \theta_{-i}) = f(\theta'_i, \theta_{-i}) = 1$ and hence
$t_i(c, \theta_{-i}) = t_i(\theta'_i, \theta_{-i})$,
since by Note \ref{not:t} $(i)$ either both sides equal 0 or
$\sum_{j \neq i}\theta_j - \frac{n-1}{n} c$.  Further, as in Case 4,
for all $j \neq i$ we have $t_j(c, \theta_{-i}) = 0$.  But the pivotal
mechanism is pay only, so for all $j \neq i$ we have
$t_j(\theta'_i, \theta_{-i}) \leq 0$. Hence
$\sum_{j=1}^{n} t_j(c, \theta_{-i}) \geq
\sum_{j=1}^{n} t_j(\theta'_i, \theta_{-i})$ and consequently (\ref{equ:social})
holds.
\II

\NI 
$(ii)$ Fix $\theta \in \Theta$. By $(i)$ and Lemma \ref{lem:optn}
it suffices to prove that for all vectors $s'_{-n}(\cdot)$ of optimal
strategies for players $1, \LL, n-1$
\[
SW(\theta, s(\cdot)) \geq SW(\theta, (s'_{-n}(\cdot), s_n(\cdot))),
\]
where $s(\cdot)$ is the vector of strategies $s_i(\cdot)$ of Theorem \ref{thm:social}.
\II

\NI
\emph{Case 1} $\sum_{j=1}^{n-1} \theta_{j} < c$.

Then by Lemma \ref{lem:compat}$(i)$ $[s(\cdot), \theta] = [(s'_{-n}(\cdot), s_n(\cdot)), \theta]$,
so in this case actually an equality holds.
\II

\NI
\emph{Case 2} $\sum_{j=1}^{n-1} \theta_{j} \geq c$.

Then $\sum_{j = 1}^{n} \theta_{j} \geq c$, and by the fact that the pivotal mechanism
is pay only, the social welfare $SW(\theta, s^{*}(\cdot))$
is maximal when the vector of players' strategies $s^{*}(\cdot)$ reduces all taxes to 0.
This is the case for $s(\cdot)$ since
in this case players $n-1$ and $n$ both submit the type $c$,
as a result of which no player is pivotal.
\HB
\II

\NI
\textbf{Proof of Theorem \ref{thm:nash}}.

\NI
$(i)$
The proof for both vectors is the same. Take a strategy $s'_i(\cdot)$ of player $i$.

\NI
\emph{Case 1} $s'_i(\cdot)$ is  not optimal.

Then $s_i(\cdot) >_d s'_i(\cdot)$ by the optimality of $s_i(\cdot)$, so
$(s_i(\cdot), s_{-i}(\cdot)) \succeq_i (s'_i(\cdot), s_{-i}(\cdot))$.
\II

\NI
\emph{Case 2} $s'_i(\cdot)$ is optimal.

Then $s_i(\cdot) =_d s'_i(\cdot)$, so we need to prove that for all $\theta \in \Theta$
\[
u_i(f([s(\cdot), \theta]), \theta_i) \geq u_i(f([s'(\cdot), \theta]), \theta_i),
\]
where $s'(\cdot) := (s'_{i}(\cdot), s_{-i}(\cdot))$.
If $i = n$, this is a direct consequence of the fact that $s_n(\cdot)$ is optimal.
Otherwise fix $\theta \in \Theta$.
\II

\NI
\emph{Subcase 1} $\sum_{j = 1}^{i}\theta_j <  c$.

Then by Lemma \ref{lem:compat}$(i)$ we have
$[s(\cdot), \theta] = [s'(\cdot), \theta]$,
so in this case actually
$
u_i(f([s(\cdot), \theta]), \theta_i) = u_i(f([s'(\cdot), \theta]), \theta_i)
$.
\II

\NI
\emph{Subcase 2} $\sum_{j = 1}^{i}\theta_j \geq c$.

Then, when each player $k$ follows the $s_k(\cdot)$ strategy,
players $i, i+1, \LL, n$ submit the type $c$, so
$f([s(\cdot), \theta]) = 1$ and all taxes are reduced to 0. As a result
$u_i(f([s(\cdot), \theta]), \theta_i) = \theta_i - \frac{c}{n}$.

Further, by Lemma \ref{lem:compat}$(iii)$ and $(iv)$ we have $f([s'(\cdot), \theta]) = 1$, so
\[
u_i(f([s'(\cdot), \theta]), \theta_i) = \theta_i - \frac{c}{n} + t_i([s'(\cdot), \theta]) \leq u_i(f([s(\cdot), \theta]), \theta_i),
\]
since the pivotal mechanism is pay only.
\II

\NI
$(ii)$ By Theorem \ref{thm:social}$(ii)$.
\HB

\end{document}

%% file: ecmds.tex
\newcommand{\oldbfe}[1]{\begin{bfseries}\emph{#1}\end{bfseries}}

\newcommand{\myra}{\mbox{$\:\rightarrow\:$}}

\newcommand{\LL}{\mbox{$\ldots$}}

\newcommand{\C}[1]{\mbox{$\{{#1}\}$}}           % curly braces

\newcommand{\NI}{\noindent}
\newcommand{\HB}{\hfill{$\Box$}}

\newcommand{\III}{\vspace{3 mm}}
\newcommand{\II}{\vspace{2 mm}}

\newcommand{\szkew}[1]{\relax \setbox0=\hbox{\kern -24pt $\displaystyle#1$\kern 0pt }%
\box0}
{\catcode`\@=11 \global\let\ifjusthvtest@=\iffalse}
\newcounter{oldmycaption}